# DRIVERS OF
# THE COST OF SPREADSHEET AUDIT

David Colver
Director, Operis Group PLC
Dcolver@operis.com

**ABSTRACT**

*A review of 75 formal audit assignments shows that the effort taken to identify defects in financial models taken from the domain of limited recourse (project) finance is uncorrelated with common measures of the physical characteristics of the spreadsheets concerned.*

## 1    THIS PAPER

Croll [1] has described a "a spreadsheet model audit process typical of that presently used in the City [ie the financial services sector] of London".  He asserted that "Discoverable model characteristics such as formula length, ratio of original to repeated cells, numbers of cell precedents and dependents and the locality and non-locality of cell linkages can be used to infer information about the relative ease or difficulty and time to review a given model."

It makes intuitive sense that a ten-cell trivium should be quicker to check than a multi-megabyte monster.  At a finer-grained level, however, this paper offers evidence that widely used metrics are in fact poor guides to the time taken to review a given model.

## 2    BACKGROUND

Project finance is an informal term for making loans on terms which\ include agreement that opportunities for recourse in the event of default will be limited to the assets being financed.  Developed initially in the natural resources industry, project finance has in the last decade been applied increasingly to the funding of infrastructure projects, such as airports, roads, bridges, power stations, prisons, hospitals and schools.  Of the projects identified by InfraNews, a news service specialising in the subject, as reaching financial close around the world in 2009, a role in over half was performed by Operis, a specialist in project finance based in London.

Decisions whether to provide more general corporate finance are typically made on the basis of financial statements.  Those essentially record what has happened in the past.  What is distinctive about project finance is the centrality in the credit evaluation process of projections of financial performance in the future.  The projections are derived from financial models.

Operis develops these financial models, both as a stand-alone service and as part of a wider remit as financial adviser, and provides training in the development of models of this kind.  It is best known for conducting formal audits of models that others have prepared.  It also sells software (OAK, the Operis Analysis Kit) relevant to these activities.

Operis is therefore exposed to the issues surrounding the auditing of spreadsheets used to price, structure and illustrate substantial transactions on a significant scale. The objective of a full formal audit is to become sufficiently confident that material defects have been removed from a spreadsheet to deliver a letter, addressed to the prospective lenders, that they can rely on the figures presented to them. The liability cover provided if such an opinion later turns out to be ill-founded can run to the tens of millions of pounds.

As part of its continuous quest for improved productivity in this activity, Operis records the effort it takes to complete these reviews in a time recording system, custom designed to track not only to which projects hours are devoted, but how they are accounted for by the different tasks on the workplan. Tasks that, over a number of assignments, consistently take longer than expected from standard measures are candidates for investigation, which might lead to improved training, partial automation of the task or process reengineering aimed at doing away with the step altogether.

## 3    DATA

Operis has analysed the data recorded on its timesheets for the audits it has recently conducted. Operis has then correlated the data with the value of the transaction, as a proxy for deal size, and various measures of the size of spreadsheet.

- How large the document is in megabytes
- How many worksheets it contains*
- How many unique formulae are involved*
- The maximum and average complexities of the formulae, measured by the number of operators and functions they contain.*

The starred items were extracted using Operis's software product, OAK v4.

The transaction values are exact for projects that were structured in Euros. They are approximate for projects involving other currencies, as they were converted to Euros at the rates prevailing at the time of writing this paper rather than the rate that actually applied when the project reached financial close; but it still gives some measure of the scale of the asset being modeled.

Also included in the database is the number of different versions, or iterations, of the spreadsheet that were reviewed. The significance of this detail is explained in section 6, Interpretation.

The database used for the analysis covers 75 assignments, concerning spreadsheets responsible for the structuring of over €54bn of financing[1] . Attention has been confined to audits completed in recent months, so that they are reasonably consistent in terms of audit process as it evolves in light of experience over time.

## 4    ADJUSTMENTS

It is necessary to exclude some samples from the database for a variety of reasons.
- Some models have been audited already, and are submitted for re-examination because they have been adjusted to reflect some change in the deal. This can happen several times in the life cycle of a transaction. The second and any subsequent inspections are artificially short as they can reuse much of the earlier work.

---

[1] €54bn is the simple total of the financings represented in each spreadsheet. A small number of transactions, including the €14bn largest, are represented by two or more models in the sample.

- Sometimes a standard template model is applied to a series of similar transactions. Again, the economics of the later audits are flattered by the ability to reuse earlier work.
- Some assignments are for a smaller scope of work than a full formal audit. Operis terms such exercises High Level Reviews.
- Some assignments ended prematurely because the initiative to which they were directed collapsed or was cancelled.

It also turns out that a fair proportion of the spreadsheets were provided in the format used by Excel 2007. This is a recent development, Excel 2007 having achieved widespread adoption only slowly. However, on the evidence of this sample, the migration is now well under way. With Excel 2007 Microsoft introduced a new file format that uses compression to store spreadsheets more compactly, a manoeuver that makes sense given that Excel 2007 permits spreadsheets to be much larger. Since Excel 2003 and 2007 file sizes are not directly comparable, all the Excel 2003 workbooks have been re-saved in the Excel 2007 xlsm format, and it is that measure that is used as an indication of file size.

## 5 ANALYSIS

The relationship between the items listed above as logged for each assignment is set out below, as measured by R-squared.

| A | B | C | D | E | F | G | | |
|---|---|---|---|---|---|---|---|---|
| 1.00 | 0.02 | 0.00 | 0.00 | 0.01 | 0.03 | 0.02 | A | Transaction value (Euros) |
|  | 1.00 | 0.00 | 0.07 | 0.01 | 0.03 | 0.01 | B | Document size (Mb) |
|  |  | 1.00 | 0.04 | 0.17 | 0.02 | 0.01 | C | Unique formulae |
|  |  |  | 1.00 | 0.27 | 0.01 | 0.08 | D | Average complexity |
|  |  |  |  | 1.00 | 0.01 | 0.04 | E | Maximum complexity |
|  |  |  |  |  | 1.00 | **0.27** | F | Number of Iterations |
|  |  |  |  |  |  | 1.00 | G | Hours logged |

All R-squareds are below 0.2 except for the ones linking

- average and maximum formula complexity: this may be dismissed as trivial as the maximum drives the average;
- number of iterations to hours logged in the audit process. Even that R-squared is only 0.27.

Proceedings of EuSpRIG 2011 Conference
"Spreadsheet Governance – Policy and Practice", ISBN: 978-0-9566256-9-4
Copyright © 2011 EuSpRIG (www.eusprig.org) and the Author(s)

A more sophisticated analysis uses a multiple regression, exploring the relationship between items A-F with item G, Hours logged. This allows for the possibility that the number of hours logged in auditing a spreadsheet is driven by some of the spreadsheet characteristics in combination. For reasons of commercial confidentiality, the number of hours consumed by each audit is represented in a normalised form, as a percentage of the average hours of all 75 assignments in the database.

|   |                      | Coefficients | Standard Error | t Stat |
|---|----------------------|--------------|----------------|--------|
|   | Intercept            | -12.78%      | 25.98%         | -0.49  |
| A | Transaction value (€m) | 0.00%      | 0.00%          | -0.37  |
| B | Size (megabytes)     | -0.13%       | 1.00%          | -0.13  |
| C | Unique formulae      | 0.00%        | 0.00%          | -1.20  |
| D | Average complexity   | 18.42%       | 9.98%          | 1.85   |
| E | Maximum complexity   | 0.35%        | 0.51%          | 0.69   |
| F | Iterations           | 6.54%        | 1.48%          | 4.41   |

This suggests that average formula complexity is on the verge of joining the number of iterations as a significant contributor to the time taken to complete an audit. To the extent this is true, we can infer that an increase of 1 in the number of terms and operators in the average formula increases the time to audit the spreadsheet by about 18%.

## 6  INTERPRETATION

A typical project finance model connects project cash flows, which describe the revenues and costs that arise from building and operating an asset, with financing cash flows, which concern how the asset construction is paid for. The modelling of the financing cash flows may well be more extensive in larger transactions, largely because it may be necessary to draw on a greater number of sources of finance to get a sizeable project financed.

The complexity of the project cash flows, by contrast, is influenced by the nature rather than the size of the project. Some have many streams of revenue or costs, built up in intricate ways, and in others the revenues and costs can be derived very simply. It is therefore no surprise that the correlations between the value of the transaction and the various measures of spreadsheet size are positive, but not especially strong.

Less obviously, the correlation between the number of formulae and the effort required to audit the model is very low.

Operis's first rationalisation of this result is that auditing a financial model does not simply involve checking a spreadsheet. The process followed by Operis, once the formalities of engagement are completed, involves a process of reviewing successive iterations of the model as they are refined by the client, in a process detailed below.

## ITERATIVE REVIEW PROCESS FOR A FULL FORMAL AUDIT

1 The client sends a spreadsheet model, along with relevant documentation such as loan agreements or subcontracts

2 Operis reviews the model, and delivers a report itemising anomalies that it has found

3 The client fixes the model and/or gets its lawyers to align the contractual paperwork with the model; in due course, it delivers a second version of model and the documents

4 Operis reviews the second version and updates its report

> *High level reviews, excluded from the database surveyed in this study because they have a more economical scope of work, stop at this point. Only the original version of the spreadsheet and maybe one revision are examined, and no formal opinion letter is delivered.*

5 The client and its lawyers fix any remaining issues and deliver a third version of the spreadsheet and documents

6 Operis reviews the third version and updates its report

and so on (for an average of 8.1 versions in the assignments studied) until:

7 Operis undertakes final quality control processes, and reviews sensitivity analysis specified by the banks and prepared by the client

8 Operis attends financial close, the legal ritual at which all contracts are signed, including the credit (loan) agreement

9 Operis delivers a letter setting out its formal opinion about the fitness for its intended purpose of the model.

This process is aimed at ensuring a separation between the teams that develop and review the spreadsheets, so that the auditor is at no point auditing his own work.

Some clients react well to the reports and address the issues raised in them quickly and effectively. Others are less skilled, and need several attempts before they get the spreadsheets right. As a result, the number of iterations of the model increases, and with it the amount of hand-holding necessary. It is this activity that is the primary driver of the cost of the review exercise. (For clarity, it is the hours expended by Operis, not the client, that are the focus of this analysis.)

At any point in this process, the deal can change as it is the subject of continuing negotiations. That too adds to the number of versions of the model that need examining, just considered in isolation. But isolation is probably inappropriate, as the phenomenon likely interacts with the skill of the model developer. Skilled modelers may be expected to handle change in the high pressure circumstances of getting a transaction concluded with more serenity, and fewer adverse consequences for the integrity of the spreadsheet, than individuals who are at the edge of their competence.

Operis's second rationalisation of the low correlation between the spreadsheet metrics presented and the effort to audit is that there are diverse ways to audit a spreadsheet. One way is to check the formulae one by one. Panko has blessed this approach repeatedly in his addresses to past Eusprig conferences, inferring from similarities between spreadsheeting and traditional software development in the nature and incidence of types of error that what has emerged as good practice in traditional software development is relevant to spreadsheets also. However, double checking Panko's seminal paper, What We Know About Spreadsheet Errors [2], shows that he gives equal weight to code inspection and to data testing, of which formula reperformance is arguably a form.

> *"Although we still have far too little knowledge of spreadsheet errors to come up with a definitive list of ways to reduce errors, the similarity of spreadsheet errors to programming errors suggests that, in general, we will have to begin adopting (yet adapting) many traditional programming disciplines to spreadsheeting".*
>
> *"In programming, we have seen from literally thousands of studies that programs will have errors in about 5% of their lines when the developer believes that he or she is finished (Panko, 2005a). A very rigorous testing stage after the development stage is needed to reduce error rates by about 80% (Panko, 2005a). Whether this is done by **data testing, line-by-line code inspection, or both**, testing is an onerous task and is difficult to do properly".*

Exhaustive formula checking is described by Croll in the paper already mentioned, and certainly used by some large accounting firms. Alternative approaches are to reconstruct the model, by:

- building an entirely new one from scratch and reconciling the outputs, as one Big Four accounting firm prefers to do;

- keying the assumptions into its own trusted, standard model, as a smaller accounting firm claims to do;

- reperforming or reconciling independently the revenue, the costs, the taxes, the debt, the equity, and the financial ratios, as Operis does.

The time taken to reconstruct the distinct modules of a spreadsheet in a systematic and much-practiced way is unrelated to the number of formulae that the spreadsheet being tested happened to use, which may be highly variable for stylistic reasons.

Reperformance has a number of practical advantages over formula checking. Insertion of a single row near the top of a worksheet has the potential to alter every formula on that worksheet. Auditors who follow the formula checking approach are therefore often led to impose limits on what can be changed in the spreadsheet. Some insist that nothing is altered in the final week before the deal is signed. Others allow changes, but only at the bottom of each worksheet: no rows are to be inserted or deleted. Reperformance and reconciliation can be done in a way that is tolerant of the repeated examination of successive, potentially changing, versions of a model described in the audit process. Given that the number of iterations inspected for each model in the sample averaged 8.1, tolerance of change has real practical value to the customer.

Checking individual formulae amounts to determining whether the route followed to derive a number is the right one by following again that same route, this time very carefully. Reperformance and reconciliation amount to pursuing a route that is intentionally different from the first one and seeing if it takes us to the same place. Operis instinctively has more comfort with the two-route approach, likening it to a surveyor's use of triangulation or an accountant's insistence on double entry.

The most important advantage of audit methods involving reperformance and reconstruction is that it is hard to maintain focus during audit by formula inspection. Panko acknowledges this issue in the sentence following the extract above:

> *"In code inspection, for instance, we know that the inspection must be done by teams rather than individuals and that there must be sharp limits for module size and for how many lines can be inspected per hour"*

To state the point more starkly, formula inspection is boring. It is difficult to retain staff who are willing to continue with it for any extended interval. This means that the average experience levels of those doing the work is likely to be lower in firms whose spreadsheet review methodologies involve that approach. A number of firms are explicit in making it a temporary rite of passage on the way to a more interesting job. Operis is doubtful that formula inspection is the One Right Way to perform spreadsheet verification on any scale, but even if it is, this retention/experience point more than offsets it in practice.

Operis's third rationalisation of the low correlation between the spreadsheet metrics presented and the effort to audit is that the hours logged are given equal weight in this analysis regardless of who contributed them. What can take a new recruit many days to complete can be done by a manager with years of experience in a few minutes. As a rule the different steps in an audit are allocated to suitable levels of experience and seniority drawn from the pool of consultants. The vagaries of scheduling can mean that some tasks get done by individuals who are arguably over-qualified for the task, in order to meet transaction deadlines.

A fourth rationalisation is that the recipients of an opinion letter vary in how clean they want that opinion to be. Some require essentially every last defect to be driven out of models before they will advance loans on the basis of them. Others prioritise getting a transaction concluded, and are more readily satisfied that the projections provided are close enough. To the extent shortfalls remain in the model, compared with the ideal, these organisations are content for them to be listed as qualifications to the reported opinion. There will tend to be more iterations of a model before the first group is ready to close a deal than are demanded by the second. One manifestation of the general risk aversion since 2007 has been the migration of institutions from the second category to the first. That aside, the tolerance for risk varies fairly randomly from institution to institution, and even among individuals responsible for transactions.

## 7      CONSEQUENCES

When asked to quote to review a financial model, Operis used to base its price on the various measures of spreadsheet size. Operis has in OAK, the Operis Analysis Kit, a product that has been engineered to perform, rapidly, this measuring function among many others. Competing firms use a similar approach to pricing financial model audits. However, Operis has now ceased to use these metrics other than in exceptional conditions, in light of the good data it has amassed showing that they don't capture the real drivers of spreadsheet review costs.

When developing, rather than auditing, a spreadsheet Operis follows a methodology that is distinctive. At Eusprig's 2010 conference, Tom Grossman compared the Operis approach with two others [3].

These approaches share common aims, and agree on much more than they disagree over. One area of divergence, though, is over the practice of staging intermediate results, that is, marshalling at the top of a block of calculation the items needed by that calculation. The benefit is that the formulae that do the most meaningful calculations can be seen to refer to cells that are very near at hand, making them quick and easy to check. The cost is that the spreadsheet is much taken up by simple formulae devoted to restating and marshalling locally to a calculation data that has been derived elsewhere in the model. To give some idea of the scale of this effect, Operis has rewritten some competitor models, which do use intermediate result staging, in its own style, which does not advocate this redundancy, and found that the result is typically as much as three times more compact.

Some followers of the staging methodologies have found themselves penalised when they come to procuring audit for the resulting models. The audit firms approached have been open that it is due to the distended formula counts. It follows from the analysis presented here that that handicap originates from an understanding of the cost of model audit that wants for sophistication, and should not be viewed as a defect of the modeling methodology (even if the methodology is one that competes for attention with Operis's own).

## 8    NO PRECEDENT

The conclusions in this paper address a highly streamlined process for reviewing on a substantial scale spreadsheets that are confined to a narrow domain. They do not necessarily extend completely to a wider range of spreadsheets of arbitrary purpose. The dispersion seen in this study could be even wider in a randomly drawn sample from arbitrary disciplines, as would be the case in a financial institution that sought to embark on a firm-wide programme of spreadsheet remediation.

Audit costs are here shown to be driven to a degree by the process of iterating spreadsheet models with the developer or owner until the process of aligning them with the deal documentation is judged complete. It is possible to imagine an internal spreadsheet audit function within a financial institution for which this iteration between model developer and auditor has no parallel or relevance. In the absence of the iteration, the spreadsheet review effort could be more correlated with common metrics of spreadsheet size. More often, however, it would seem likely that a discovery of defects in spreadsheets would be followed up by a process of addressing the issue, which will likely involve human interactions of some kind that are roughly analogous to the formal iteration described.